\documentclass[intlimits,twoside,a4paper]{article}

\usepackage{amsmath,amssymb}
\usepackage{graphicx}
\usepackage{color}

\usepackage[T2A]{fontenc}
\usepackage[cp1251]{inputenc}

\usepackage[eqsecnum]{cmpj2}

\issue{2016}{19}{2}{23602}
\doinumber{10.5488/CMP.19.23602}

\title[The effect of stirring on the heterogeneous nucleation of water\ldots]{The effect of stirring on the heterogeneous nucleation of water and of clathrates of tetrahydrofuran/water mixtures
}
\author[P.W. Wilson, A.D.J. Haymet]{P.W. Wilson\refaddr{1}\thanks{E-mail: Peter.w.wilson@utas.edu.au}\,, A.D.J. Haymet\refaddr{2}}
\addresses{
\addr{1} Faculty of Science, Engineering and Technology, University of Tasmania, Hobart 7001,  Australia
\addr{2} Scripps Institution of Oceanography, University of California, San Diego, \\ 9500 Gilman Drive, San Diego, CA 92093, USA
}

\date{Received November 20, 2015}
\authorcopyright{P.W. Wilson, A.D.J. Haymet, 2016}

\begin{document}

\maketitle

\begin{abstract}
The statistics of liquid-to-crystal nucleation are measured for
both water and for clathrate-forming mixtures of tetrahydrofuran
(THF) and water using an automatic lag time apparatus (ALTA). We
measure the nucleation temperature using this apparatus in which a
single sample is repeatedly cooled, nucleated and thawed.  The
effect of stirring on nucleation has been evaluated numerically
and is discussed.  We find that stirring of the solution makes no
difference to the nucleation temperature of a given solution in a
given tube.
\pacs  64.60.Q-
\end{abstract}

\section{Introduction}

We have measured the heterogeneous nucleation of both water and a clathrate-forming liquid mixture, tetrahydrofuran (THF)/water in order to determine if stirring the solution has any effect on the nucleation temperature.  All measurements were made at one atmosphere pressure.

Tetrahydrofuran (THF) C$_4$H$_8$O is also known as butylene oxide, 1,4-epoxybutane, cyclotetramethylene oxide, oxacyclopentane, diethylene oxide, oxolane, and furanidine.  With water it is known to form Structure II clathrates, at one atmosphere, below a temperature which depends on concentration.  For a THF/water mixture at a concentration of 81 mass \% water (a 17:1 mole ratio), the melting point of the clathrate is raised to approximately $+4.5^\circ$C.

The lag-time before a supercooled sample nucleates to a solid is a stochastic quantity, strongly dependent on the degree of supercooling.  In chemical engineering, and some other fields, it is called the ``induction'' time \cite{1}.

Our experimental procedure involves taking a small volume of
solution and cooling it linearly to below its equilibrium melting
point until it nucleates and freezes.  The time for the actual
freezing to occur is very, very short compared to the cooling time
and is not taken into account in any further analysis below.  A
supercooled solution will often spontaneously freeze when bumped
or tapped.  What is less well known is any possible effect that
stirring of the solution may have on nucleation.  Stirring was
also designed to mimic turbulence in oil pipelines, where the
nucleation of hydrates is a flow assurance problem.

\section{Materials and methods}

The measurements described here are made on a purpose built automated nucleation device that we have termed automatic lag time apparatus (ALTA) \cite{2}  The water used is Ultrapur reagent grade water (Merck, Germany) filtered through a 0.2~{\textmu}m filter.  The THF is reagent grade from Scharlau, Spain.

\begin{figure}[!t]
\centerline{
\includegraphics[width=0.75\textwidth]{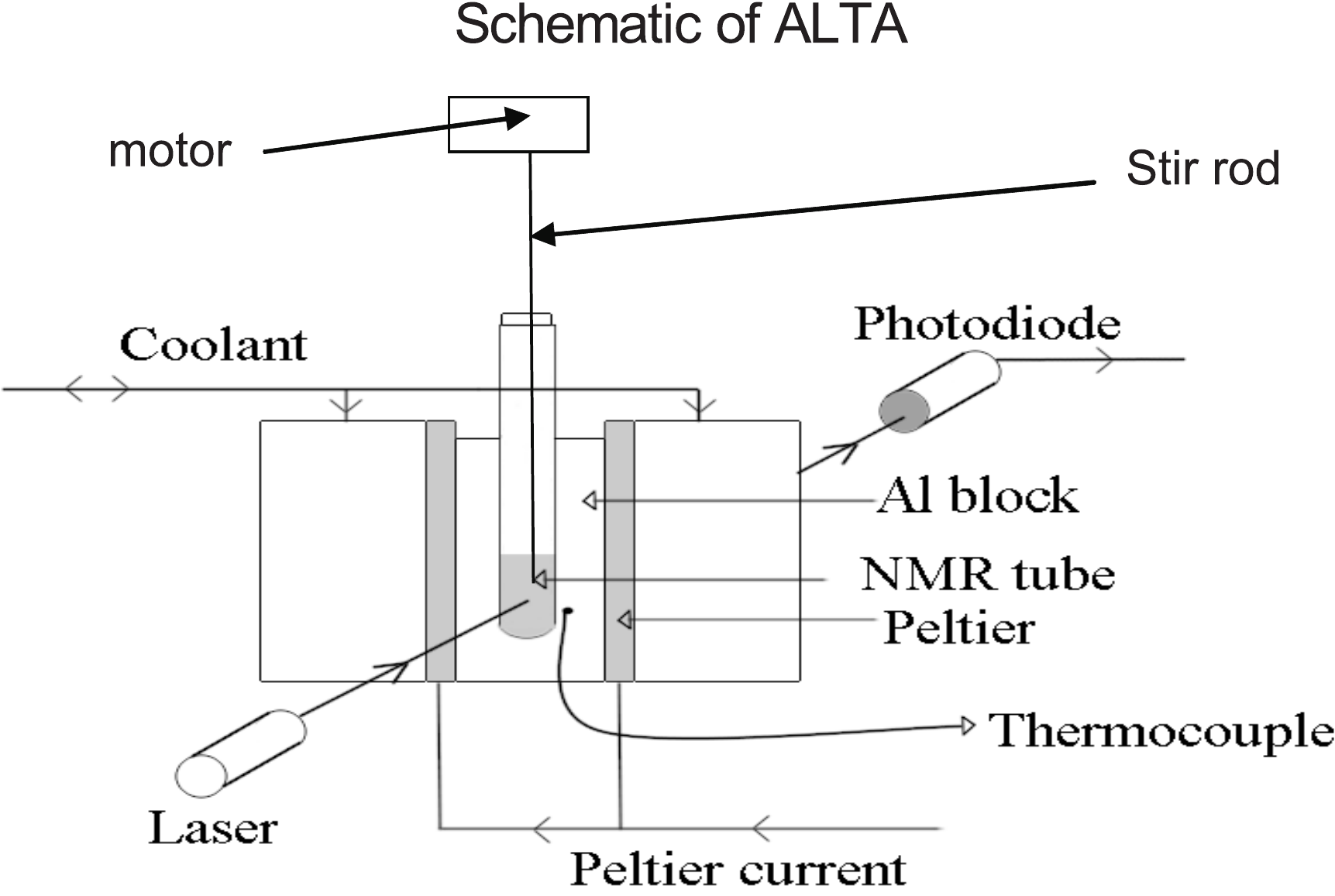}
}
\caption{The experimental arrangement of ALTA.}
\label{fig1}
\end{figure}

Sample volumes of 300~{\textmu}l have been used in each case and
are placed in purpose built sample tubes made from borosilicate
glass with an outside diameter of 5.0~mm and a length of 65~mm.
These glass tubes are inserted with a snug fit into a hole drilled
into an aluminium sample holder in the ALTA (see figure~\ref{fig1}).  The aluminium block
is cooled by thermoelectric units on each side and the temperature
control is by a PID package built into a Genie software package
(Advantech Inc.) which controls the experiment via a multipurpose
DAQ card and PC interfacing.  A cooling rate of 4.5~K~min.$^{-1}$
is used and freezing of the samples is detected optically due to a
sudden lowering in the optical transmission of a laser beam passed
through the sample tube. Once freezing has been detected, the
software causes heating of the tube for 300~s to ensure a complete
melting of any ice or hydrates, before the same tube and sample is
cooled again.  This cycle is repeated more than 300 times in any
single measurement run to gather statistically valid data about
the nucleation temperature of the sample in that tube.

\begin{figure}[!b]
\centerline{
\includegraphics[width=0.5\textwidth]{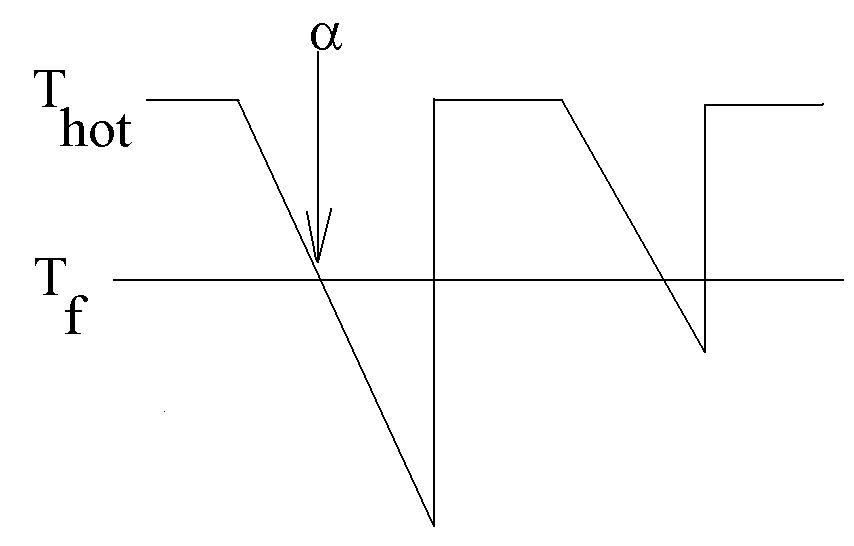}
}
\caption{The experimental protocol of cooling linearly and heating and holding, once freezing has been detected during each cycle.}
\label{fig2}
\end{figure}

A glass capillary (2~mm diameter) with a scalloped lower section
of 2~mm length was also inserted in the sample tube for every
measurement. For some series of runs (usually 50 freeze/thaw
cycles), it was stationary and then rotated around the vertical
axis at either 600 (slow) or 2000 (fast) revolutions per minute. A
motor speed controller was used to control a small DC motor
connected to the stir rod, which passed through a bush in the top
of the box of ALTA, which is flushed with dry nitrogen in order to
avoid condensation on all the cold parts.

Figure~\ref{fig2} shows the experimental protocol for the ramping down of the sample temperature until freezing occurs.  This cycle is repeated more than 300 times in any single measurement run to gather statistically valid data about the nucleation temperature of the sample in that tube.

\section{Results and discussion}

\begin{figure}[!t]
\centerline{
\includegraphics[width=0.65\textwidth]{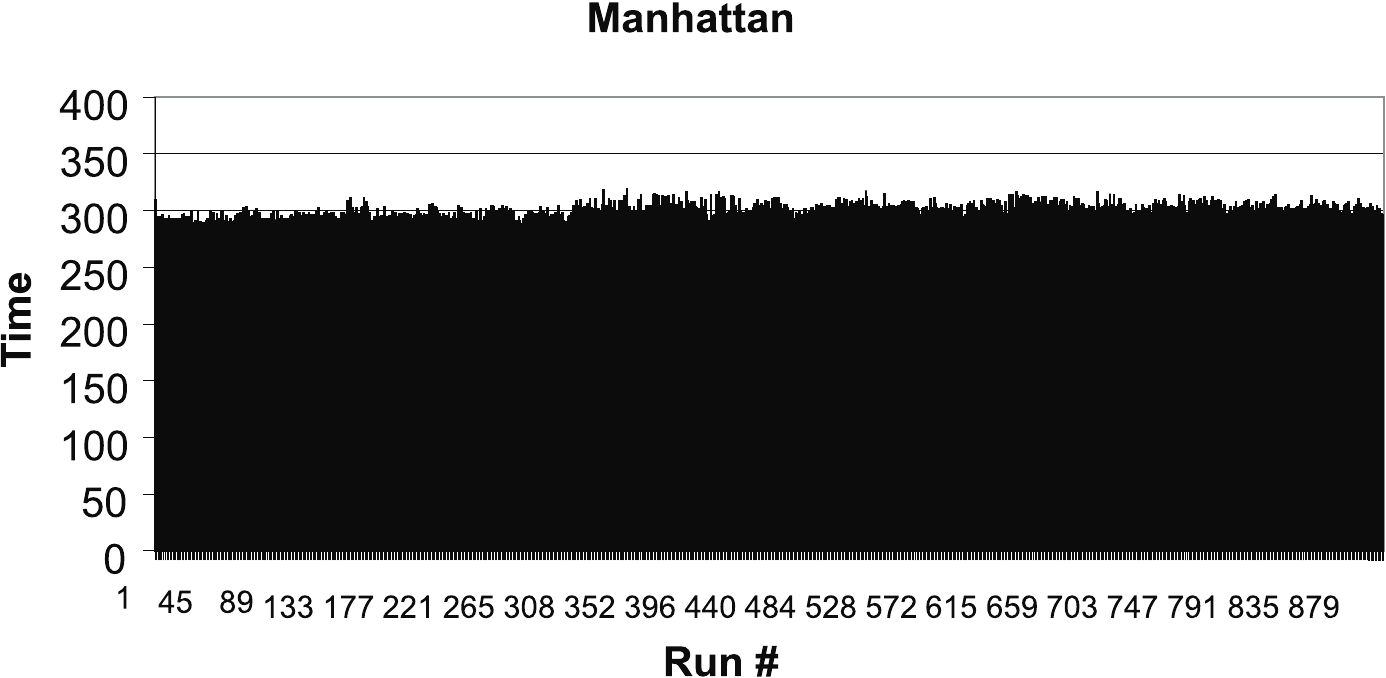}
}
\caption{Typical ALTA raw data, in the form of a Manhattan.}
\label{fig3}
\end{figure}

The data collected are the lag-time $\tau$ until nucleation and
the super cooled temperature $\Delta$T at which nucleation occurs
for each run.  When the time for each successive run is plotted as
a histogram, we produce a data set such as that shown in
figure~\ref{fig3}, usually known as a Manhattan~\cite{3}.

\begin{figure}[!b]
\centerline{
\includegraphics[width=0.65\textwidth]{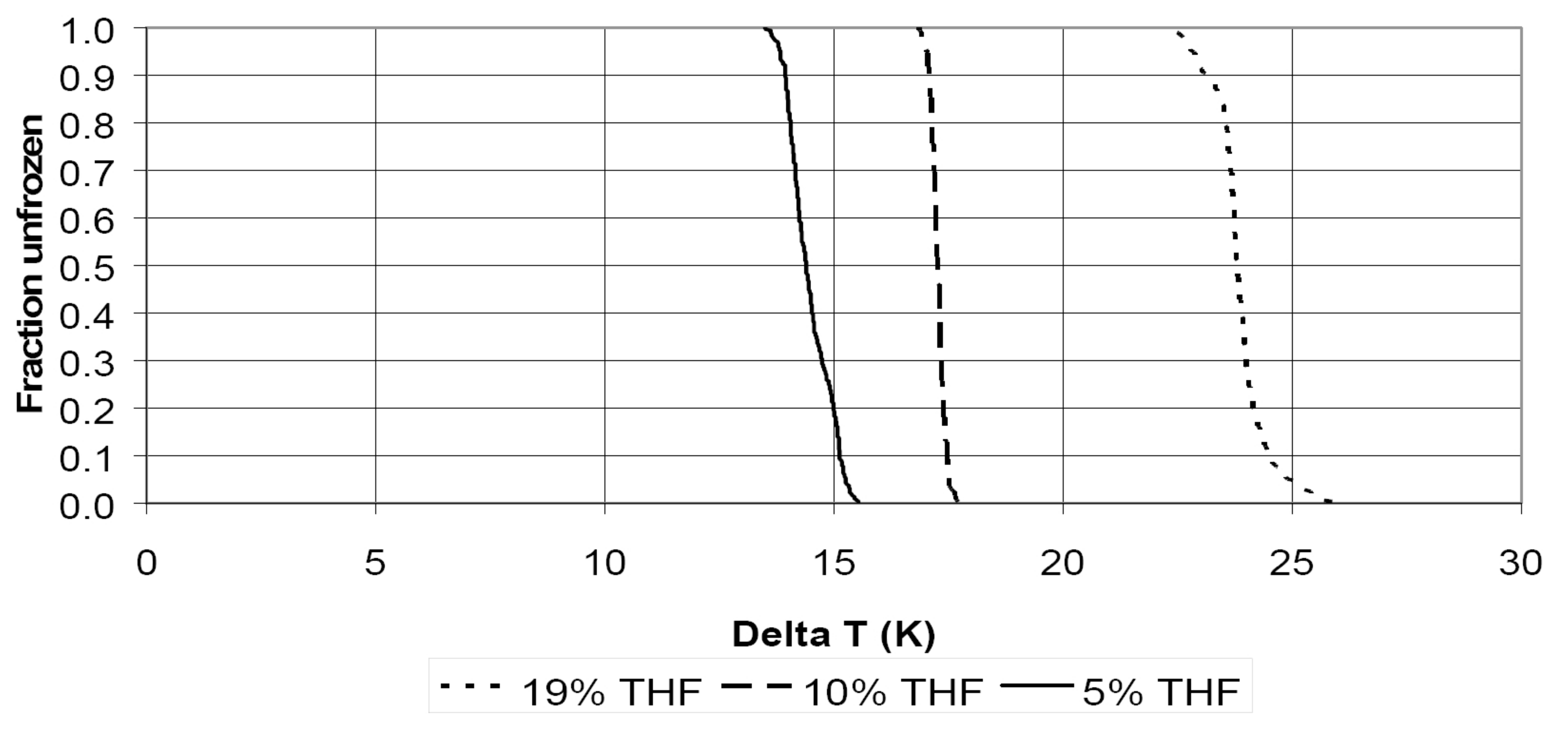}
} \caption{The survival curves for three runs, each at a different
THF concentration, are shown here and are typical of any run at
that concentration, in that tube.  Note also that the $x$-axis is
the level of supercooling, $\Delta T$, and not the actual
temperature in $^\circ$C.  This is because the melting points of
the various concentrations of THF vary.  We plot the levels of
supercooling in this way in order to compare the effects of
stirring on the T50.} \label{fig4}
\end{figure}

We now define the `survival curve' as the number of samples
unfrozen after time $t$, $N(t)$ divided by the total number of
samples $N_0$.   The survival curves for a typical set of three
runs, is shown in figure~\ref{fig4}.  The $y$-axis is the fraction
of runs as yet unfrozen at that temperature while at colder
temperatures all of the runs had frozen.  We are defining the
T50\% as the kinetic freezing point (also known as the
supercooling point), i.e., the temperature at which half runs on
that sample are still unfrozen.  The $y$-axis is the fraction
unfrozen at a given temperature.  Clearly, at warmer temperatures
all of the runs were as yet unfrozen while at colder temperatures
all of the runs had frozen.  Note also that the $x$-axis in now
level of supercooling, $\Delta T$, and not the actual temperature
in $^\circ$C.  This is because the melting points of the various
concentrations of THF vary.  We plot the levels of supercooling in
this way in order to compare the effects of both concentration and
the added substances or changed conditions (such as stirring) on
the nucleation temperatures.  The curves in figure~\ref{fig4} are
typical of all the curves we have generated for each series at
each concentration.  For the three data sets shown in
figure~\ref{fig4}, the T50s are 14, 17 and 23.5 degrees of
supercooling for the 5, 10 and 19~wt~\% THF concentrations,
respectively.

\begin{table}[!t]
\caption{T50 points for at least 300 runs each, with water and THF/water under various stirring regimes.}
\label{tab}
\begin{center}
\begin{tabular}{|c||c|c|c|c|}
\hline\hline
Sample  & No stir bar & No stirring & Slow stirring & Fast stirring \\ \hline\hline
water & $-9.9$ & $-10.8$ & $-10.8$ & $-10.7$ \\ \hline
water && $-9.4\phantom{0}$  &$-9.6\phantom{0}$  & \\ \hline
water  && $-12.0$ & $-11.0$  & \\ \hline
THF (19\%) &&  $-14.4$ & $-14.0$  & \\
\hline \hline
\end{tabular}
\end{center}
\end{table}

\section{Conclusions}

Our experiments have shown that THF/water mixtures behave
similarly, in general, to pure water solutions.  We find that THF
at 19\% concentration, stirred at either 600 or 2000 r.p.m. does
not markedly affect the nucleation temperature in either
direction.  Thus, we conclude that turbulence probably does not
affect the nucleation temperature in pipelines, but more likely it
will be insoluble particles acting as nucleation sites.

\ukrainianpart

\title{Ефект перемішування на гетерогенну нуклеацію води \\ і клатратів
з сумішей тетрагідрофуран/вода}
\author{П.В. Вільсон\refaddr{1}, А.Д.Дж. Геймет\refaddr{2}}
\addresses{
\addr{1} Факультет науки, інженерії і технології, Університет Тасманії,
Гобарт 7001, Австралія
\addr{2} Інститут океанографії Скріппса, Університет Каліфорнії Сан-Дієго, Сан-Дієго, Каліфорнія 92093-0210, США
}

\makeukrtitle

\begin{abstract}
\tolerance=3000%
Вимірюється статистика нуклеації рідина-до-кристалу для води і для граткоформуючих
сумішей тетрагідрофурану та води, використовуючи апарат автоматичної часової затримки
(ALTA). Ми вимірюємо температуру нуклеації, використовуючи цей апарат, в якому той
самий зразок повторювально охолоджується, проходить нуклеацію і розплавлюється.
Ефект перемішування на нуклеацію був оцінений чисельно і дискутується. Ми отримали,
що перемішування не дає ефекту для температури нуклеації даного розчину в даній
комірці.

\keywords гідрати, тетрагідрофуран, нуклеація, апарат ALTA, перемішування, вода

\end{abstract}

\end{document}